\documentclass{PoS}
\PoS{PoS(LAT2005)342}

\renewcommand{\>}{\rangle}
\newcommand{\<}{\langle}
\newcommand{\beq}{\begin{equation}}
\newcommand{\eeq}{\end{equation}}
\newcommand{\beqn}{\begin{eqnarray}}
\newcommand{\eeqn}{\end{eqnarray}}

\newcommand{\half}{\frac{1}{2}}

\title{Four quark operators in maximally twisted Wilson LQCD }
                                                                                                
\ShortTitle{Four quark operators in maximally twisted Wilson LQCD }
                                                                                                
\author{\speaker{Roberto Frezzotti}\thanks{From September 1st, 2005, 
new affiliation and address: Universit\`a di Roma ``Tor Vergata'', 
Dipartimento di Fisica,
Via della Ricerca Scientifica -- 00133 Roma (Italy); e-mail: 
Roberto.Frezzotti@roma2.infn.it~.}\\
        INFN, Sezione di Milano and Universit\`a di Milano ``Bicocca'',
P.zza della Scienza 3 -- 20126 Milano (Italy) \\
                                                                                                
        \mbox{E-mail: \email{ 
                       Roberto.Frezzotti@mib.infn.it } } }
                                                                                                
\author{Giancarlo Rossi\\
                                                                                                
        Universit\`a di Roma ``Tor Vergata'', Dipartimento di Fisica,
Via della Ricerca Scientifica -- 00133 Roma (Italy) \\
                                                                                                
        E-mail: \email{Giancarlo.Rossi@roma2.infn.it}}

\abstract{We discuss how the peculiar properties of maximally twisted
Wilson fermions can be exploited to set up a consistent LQCD computational
scheme in which the CP-conserving matrix elements of the $\Delta S =1,2$
effective weak Hamiltonian can be evaluated without mixing with wrong-chirality
and/or -parity operators. The proposed lattice framework guarantees 
automatic O($a$) improvement and positive determinant also for
pairs of non-degenerate quarks.}

\FullConference{XXIIIrd International Symposium on Lattice Field Theory\\
                                                                                                
                 25-30 July 2005\\
                                                                                                
                 Trinity College, Dublin, Ireland}

\begin{document}

\section{Introduction and main results}

We outline here a rather general strategy~\cite{FR2} to compute
weak matrix elements --and more generally operator matrix elements 
among hadronic states-- using {\em maximally twisted} Wilson fermions
in such a way that wrong-chirality and -parity mixings, as well
as O($a$) discretization errors, are avoided and positivity of
the fermionic determinant is guaranteed~\footnote{We do not discuss
here the various possible ways of performing the (non-perturbative) 
renormalization of the matrix elements of interest. However,
if one wishes to employ a Schr\"odinger functional scheme without
reintroducing O($a$) bulk discretization errors, modifications
of the usual setup are needed: see Refs.~\cite{newSF} for two different
proposals.}. Our strategy relies on
\begin{itemize}

\item  the existence of renormalizable Euclidean lattice gauge
models including ``sea'' quarks as well as possibly replicated
``valence'' quarks (and corresponding ghost fields), from which
one can extract, among others, operator matrix elements 
that in the continuum limit coincide with those of the
effective weak Hamiltonian in QCD;

\item  the possibility of choosing the (sign of the) Wilson parameters 
for the various replica of (maximally twisted, single flavour) ``valence'' 
quark fields, as well as the way they enter the relevant quark operators,
%the multilocal operators of which one needs 
%to compute the vacuum expectation value, 
so as to avoid wrong-chirality or -parity mixings. 

\end{itemize}
The latter property has so far~\cite{FR2} been established only for the 
CP-conserving $\Delta S=1,2$ effective weak Hamiltonian operators
(see discussion below) but is likely to be more general. We also 
assume in the following that the ``infrared'' problems in the lattice 
computation of matrix elements for multiparticle states~\cite{MT}
%(final state interactions in an Euclidean framework)
have been solved in some way, e.g.\ by the methods of Refs.~\cite{LL+LMST}.

We focus here on matrix elements of four quark operators. The evaluation
of matrix elements of two quark operators without unnecessary mixings and
O($a$) cutoff effects is in several cases feasible within the
framework of plain twisted mass Lattice QCD (tmLQCD)~\cite{FGSW}
(even in the case of mass  non-degenerate quark pairs~\cite{CAIRNS03})
and will not be discussed here~\footnote{We only recall a recently proposed application
of tmLQCD to the evaluation of $V_{us}$~\cite{IMT}.}.
For recent reviews on tmLQCD see Refs.~\cite{REV_TM}. 
An extension of our approach to the computation of $B_B$ in the static 
heavy quark limit has been proposed in Ref.~\cite{DM04+}.

%An extension of our approach to the computation of $B_B$ in the static 
%heavy quark limit has been proposed in Ref.~\cite{DM04+}. The computation
%of matrix elements of two quark operators with neither mixings nor
%O($a$) cutoff effects is in several cases feasible within the 
%framework of plain twisted mass Lattice QCD (tmLQCD)~\cite{FGSW} 
%(possibly with mass  non-degenerate
%quark pairs~\cite{CAIRNS03})~\footnote{For a recently proposed application
%to the evaluation of $V_{us}$ see Ref.~\cite{IMT}.}  %%% CONTROLLA %%%
%and will not be discussed here. For recent reviews on tmLQCD see 
%Refs.~\cite{REV_TM}. 

Other methods for the evaluation of specific four fermion matrix elements using twisted 
Wilson fermions with no (or reduced) wrong-chirality and -parity mixings
are possible and have in fact been developed for the cases of
$B_K$~\cite{FGSW,TMBK_DAT}, the $\Delta I =1/2$ rule~\cite{PSV04}
and $B_B$ in the static heavy quark limit~\cite{PPPW05}. 
For recent reviews on weak matrix elements in LQCD see e.g.\ 
Refs.~\cite{REV_WME}.

\section{4sNv Euclidean lattice models}                                                                                                
Let us start by specifying the lattice formulation of a
generic (Euclidean SU(3)) gauge model with ``sea'' quarks
as well as $N$ flavours of ``valence'' quarks and corresponding ghosts 
we alluded to above. For definiteness we specialize to the
realistic case of four non-degenerate physical quark flavours ($u$,
$d$, $s$ and $c$). The lattice action takes thus the form
\beq
S_{4sNv} = S_{g}[U] 
+ S_{\rm tm}^{(\pi/2)}[\psi_\ell, \bar\psi_\ell,U]
+ S_{\rm tm}^{(\pi/2)}[\psi_h, \bar\psi_h,U] 
+ \sum_{f=1}^N \Big{[} S_{\rm OS}^{(\pi/2)}[q_f,\bar{q}_f,U] +
S_{gh}^{(\pi/2)}[\phi_f,U] \Big{]} \label{FULACT}
\eeq
where~\footnote{For undefined notations here and in the following 
we refer the reader to Ref.~\cite{FR2}.} 
$S_{g}[U]$ is a suitable pure gauge
action~\footnote{In unquenched computations
a clever choice of $S_g[U]$ can be important for the phase structure 
of the lattice model and thus crucial to be able to perform simulations at maximal
twist with no metastability problems (see Refs.~\cite{MONTV04,REV_TM}).},
\beq
S_{\rm tm}^{(\pi/2)}[\psi_p,\bar{\psi}_p,U] = a^4 \sum_x\,\bar
\psi_p(x) \Big{[}\gamma\cdot\widetilde\nabla -i\gamma_5\tau_1
W_{\rm{cr}}(r_p)+m_{p} - \epsilon_{p}\tau_3
\Big{]}\psi_p(x) \, , \qquad p=\ell, h
\label{SEAQACT}
\eeq
is the action for a pair of ``sea'' quarks ($\psi_\ell = (u,d)$,
$\psi_h = (s,c)$ and typically $|r_p|=1$) with
\beq
W_{\rm{cr}}(r) = -a\frac{r}{2}\sum_\mu\nabla^\star_\mu\nabla_\mu+
M_{\rm{cr}}(r)\; ,
\label{WDONDM}
\eeq
while
\beqn
& S_{\rm{OS}}^{(\pi/2)}[q_f,\bar q_f,U] = &
a^4 \sum_x\,\bar q_f(x) \Big{[}\gamma\cdot\widetilde\nabla
-i\gamma_5 W_{\rm{cr}}(r_f)+m_f\Big{]}q_f(x) \; ,
\label{VALQACT}  \\
& S_{gh}^{(\pi/2)}[\phi_f,U] = &
a^4 \sum_x\, \phi^\dagger_f(x)
\, {\rm sign}(m_f) \Big{[}\gamma\cdot\widetilde\nabla -i\gamma_5
W_{\rm{cr}}(r_f)+m_f\Big{]}\phi_f(x) \label{VALGACT}
\eeqn
represent the action for a single flavour, $q_f$, \"Osterwalder--Seiler (OS)
fermion~\cite{OS} and that for the corresponding ghost field, $\phi_f$, 
respectively. The latter is a c-number, spin 1/2 lattice field belonging to 
the fundamental representation of the gauge group SU(3). Integration over
$\phi_f$ 
%%($\phi_f^\dagger$ merely represents its complex conjugate) 
is convergent and yields an inverse single-flavour determinant that cancels
the corresponding determinant arising from ``integration'' over the
Grassmann fields $q_f$ and $\bar{q}_f$. 

Renormalizability of the lattice 4sNv model with action~(\ref{FULACT}) follows
from standard power counting and symmetry~\cite{FR2,SHSH} arguments: for
more details, such as the precise definition of the critical masses appearing
in eqs.~(\ref{SEAQACT}) to~(\ref{VALGACT}), see Ref.~\cite{FR2}. Here we only 
recall that valence flavour is obviously conserved and all valence (current) 
quark masses renormalize multiplicatively
\beq
\hat{m}_f = Z_m(r_f) m_f \; , \label{renvalM}
\eeq
while the renormalized sea (current) quark masses take the form 
\beqn
m_{u_{\rm sea}} \equiv m_\ell^{-} = 
Z_P^{-1}(r_\ell) m_\ell - Z_S^{-1}(r_\ell) \epsilon_\ell \, ,
\quad &
m_{d_{\rm sea}} \equiv m_\ell^{+} = 
Z_P^{-1}(r_\ell) m_\ell + Z_S^{-1}(r_\ell) \epsilon_\ell \, ,
\nonumber \\
m_{s_{\rm sea}} \equiv m_h^{-} = 
Z_P^{-1}(r_h) m_h - Z_S^{-1}(r_h) \epsilon_h \, ,
\quad &
m_{c_{\rm sea}} \equiv m_h^{+} =
Z_P^{-1}(r_h) m_h + Z_S^{-1}(r_h) \epsilon_h \, .
\label{renseaM}
\eeqn
One can show that
in a mass independent scheme all these (sea and valence) renormalized quark 
masses have a common scale dependence.

In the following we will restrict attention to vacuum expectation
values of multilocal operators made out exclusively of valence quark fields.
%% Ignoring logarithmic dependencies on $a$,
Automatic O($a$) improvement (and absence of all discretization errors
coming with odd powers of the lattice spacing) of parity-even correlators
and derived quantities follows from the invariance
of the lattice 4sNv model under ${\cal P} \times {\cal D}_d \times 
(M \to - M)$, with $M$ an array made out of the $4+N$ bare quark
mass parameters. The precise definitions of ${\cal P}$ (the physical
parity operation) and ${\cal D}_d$ can be found in Ref.~\cite{FR2}. 
Since the effect of ${\cal D}_d \times (M \to - M)$ on a generic 
operator of naive dimension $d$ is to multiply it by $(-1)^d$, 
besides reflecting all its spacetime arguments, a simple argument for
automatic O($a$) improvement can be made that is closely analogous to 
the argument given in
Ref.~\cite{FMPR} for plain tmLQCD~\footnote{A slightly different argument,
which employs the $r$-parity properties of the critical masses,
was given in Ref.~\cite{FR2}.}.

\section{$B_K$ and $\Delta I=1/2$ rule: prescription for bare matrix elements}

The general strategy of our approach is as follows. 
Let $\< \beta | {\cal O} | \alpha \>$
be the matrix element of the local operator ${\cal O}$ between hadronic states
$| \alpha \>$ and $| \beta \>$ --evaluated in QCD with four flavours (qcd4)- 
we are interested in. Usually $\< \beta | {\cal O} | \alpha \>$ 
is extracted from a suitable qcd4 correlator, but, as argued
below, it can equally well be extracted, up to discretization errors,
from correlators of the form 
\beq
C_{\Phi_\beta {\cal Q} \Phi_\alpha }^{(4sNv)}(x,y) = 
\< \Phi_\beta(x_1,x_2,\dots) \, {\cal Q}(0) \,
\Phi_\alpha(y_1,y_2,\dots) \>^{(4sNv)} \, , 
\label{GENCORR}  \eeq
evaluated in an appropriate 4sNv lattice model where ${\cal Q}$, 
$\Phi_\alpha$ and $\Phi_\beta$ are suitable operators 
whose form is determined by
${\cal O}$ and the states $| \alpha \>$ and $| \beta \>$, respectively.
Our approach is illustrated in~\cite{FR2}, where we discuss in some
detail a few cases of phenomenological relevance. The main points 
of that discussion are summarized below.
One can check that in all these cases 
the number of valence (Mtm-Wilson) quark propagators to be computed
is only 1.5 times larger than usually.
%in an ordinary setup where the
%(Wilson or tm-Wilson) valence quarks are not replicated. 
%and the operators
%take their usual form (but nasty operator mixings inevitably arise).

\subsection{$K^0$--$\bar{K}^0$ mixing amplitude} \label{3A}

The target qcd4 matrix element (renormalized at the scale $\mu$), namely 
$~\langle
\bar{K}^0 | \hat{\cal O}_{VV+AA}^{\Delta S=2} (\mu) | K^0\rangle \equiv
{16 \over 3} M_K^2 F_K^2  \hat{B}_K(\mu)~$ , 
is usually extracted from the correlator ($x_0\!>\!0$, $y_0\!<\!0$)
\beq C_{K{\cal O}K}^{({\rm qcd}4)}(x,y) = \langle (\bar{d}\gamma_5
s)(x) \widehat{{\cal O}}_{VV+AA}^{\Delta S=2}(0) (\bar{d}\gamma_5
s)(y) \rangle \,
 \label{CORRBK}
\eeq
with the bare expression of 
$\widehat{{\cal O}}_{VV+AA}^{\Delta S=2}$ 
given by
$~{\cal
O}_{VV+AA}^{\Delta S=2} = (\bar{s} \gamma_\mu d) (\bar{s}
\gamma_\mu d) + (\bar{s} \gamma_\mu\gamma_5 d) (\bar{s}
\gamma_\mu\gamma_5 d)~$.
% 
%\vspace{2mm}
We propose to consider a 4s6v model, with valence quarks $u$,
$d$, $d'$, $s$, $s'$, $c$, and extract the bare matrix element
\beq
\langle \bar{K'}^0 | \, {\cal Q}_{VV+AA}^{\Delta S=2} | K^0
\rangle = {16 \over 3} M_{K'} F_{K'} M_K F_K  B_K^{\rm bare} \, .
\label{BKDEFLAT}
\eeq
from a correlator of the form~(\ref{GENCORR}) with 
$\Phi_\beta = \bar{d}' \gamma_5 s'$, $\Phi_\alpha = \bar{d} \gamma_5 s$,
\beq
 {\cal Q}_{VV+AA}^{\Delta S=2} = 2 \Big{[}
(\bar{s} \gamma_\mu d) (\bar{s}' \gamma_\mu d')
+ (\bar{s} \gamma_\mu\gamma_5 d) (\bar{s}' \gamma_\mu\gamma_5 d') +
(\bar{s} \gamma_\mu d') (\bar{s}' \gamma_\mu d)
+ (\bar{s} \gamma_\mu\gamma_5 d') (\bar{s}' \gamma_\mu\gamma_5 d) \Big{]} \, 
\label{OP_BK}
\eeq
and the lattice regularization fixed by~\footnote{The $u$ and $c$ valence flavours
do not enter explicitly here, their regularization can thus be left unspecified.}
$~r_d = r_{d'} = r_s = -r_{s'}~$. Direct symmetry arguments
thus imply~\cite{FR2} that ${\cal Q}_{VV+AA}^{\Delta S=2}$, eq.~(\ref{OP_BK}), 
renormalizes multiplicatively.

\subsection{$K \to \pi\pi$ amplitudes} \label{3B}

The amplitudes for weak $K \to \pi\pi$ decays in the $\Delta I =1/2, 3/2$ 
channels can be expressed in terms of Wilson coefficients
and the (renormalized) matrix elements 
$\< \pi^+\pi^- | \widehat{\cal O}^{\; \pm}_{VA+AV}(\mu) | K^0 \>$,
$\< \pi^0\pi^0 | \widehat{\cal O}^{\; \pm}_{VA+AV}(\mu) | K^0 \>$
%\beq
%\< \pi^+\pi^- | \widehat{\cal O}^{\; \pm}_{VA+AV}(\mu) | K^0 \> \, ,
%\qquad \< \pi^0\pi^0 | \widehat{\cal O}^{\; \pm}_{VA+AV}(\mu) | K^0 \> \, ,
%\eeq
with the bare expression of $\widehat{\cal O}^{\; \pm}_{VA+AV}(\mu)$
given by 
\beqn
{\cal
O}^{\pm}_{VA+AV} &=& \half \Big[ (\bar{s}\gamma_\mu u) (\bar{u}
\gamma_\mu\gamma_5 d) \pm (\bar{s}\gamma_\mu d) (\bar{u}
\gamma_\mu\gamma_5 u) \Big]  -  \half \Big[ u \leftrightarrow c
\Big] +
\nonumber \\
&&+ \half \Big[ (\bar{s}\gamma_\mu\gamma_5 u) (\bar{u} \gamma_\mu
d) \pm (\bar{s}\gamma_\mu\gamma_5 d) (\bar{u} \gamma_\mu u) \Big]
 -  \half \Big[ u \leftrightarrow c \Big] \, . \qquad
\label{O_pm_VAAV}
\eeqn
We propose to consider a 4s10v model, with valence quarks $u$, $u'$,
$u''$, $u'''$, $d$, $s$, $c$, $c'$, $c''$, $c'''$, and extract the
bare matrix elements~\footnote{As argued below, the renormalized
counterparts of these matrix elements yield lattice estimates of
the correspondingly renormalized qcd4 target matrix elements, 
$\< \pi^+\pi^- | \widehat{\cal O}^{\; \pm}_{VA+AV}(\mu) | K^0 \>$ and
$\< \pi^0\pi^0 | \widehat{\cal O}^{\; \pm}_{VA+AV}(\mu) | K^0 \>$.}  
$\< \pi^+\pi^- | {\cal Q}^{\; \pm}_{VA+AV} | K^0 \>$ and
$\< \pi^0\pi^0 | {\cal Q}^{\; \pm}_{VA+AV} | K^0 \>$
from (connected) correlators of the form~(\ref{GENCORR}), namely
\beqn
C_{\pm ,K^0\pi^+\pi^-}^{(4s10v)}(x_1,x_2,y) &=& \langle
 \Phi_{\pi^+\pi^-}(x_1,x_2)
{\cal Q}^{\pm}_{VA+AV}(0) \Phi^\dagger_{K^0}(y)
\rangle_{\rm conn}^{(4s10v)} \, ,
\label{K2PIC_modcorr} \\
C_{\pm , K^0\pi^0\pi^0}^{(4s10v)}(x_1,x_2,y) &=& \langle 
 \Phi_{\pi^0\pi^0}(x_1,x_2)
{\cal Q}^{\pm}_{VA+AV}(0) \Phi^\dagger_{K^0}(y)
\rangle_{\rm conn}^{(4s10v)}  \, .\label{K2PIN_modcorr}
\eeqn
The meson interpolating fields $\Phi^\dagger_{K^0}$,
$\Phi_{\pi^+\pi^-}$ and $\Phi_{\pi^0\pi^0}$ are made out
of valence $u$, $d$ and $s$ quarks only and the four fermion operator
takes the form (see eq.~(\ref{O_pm_VAAV})) 
\beq
{\cal Q}^{\pm}_{VA+AV} = {\cal O}^{\pm \, [0]}_{VA+AV}
 + {\cal O}^{\pm \, [1]}_{VA+AV}
 - {1 \over 2} {\cal O}^{\pm \, [2]}_{VA+AV}
 - {1 \over 2} {\cal O}^{\pm \, [3]}_{VA+AV} \, ,
\label{Q_01_23_COMB}
\eeq
with ${\cal O}^{\pm \, [0]}_{VA+AV}$ involving $u$ and $c$ valence
quark fields, ${\cal O}^{\pm \, [1]}_{VA+AV}$ involving $u'$ and $c'$
and so on. The key point to avoid wrong-chirality or -parity mixings
of ${\cal Q}^{\pm}_{VA+AV}$ is to fix
the lattice regularization by
\vspace{-0.2cm}
\beq
r_s = r_d = 
r_u = -r_{u'} = r_{u''} =  -r_{u'''} = r_c = -r_{c'} = r_{c''} =
-r_{c'''} \, .
\eeq

\subsection{$K \to \pi$ amplitudes} \label{3C}

Since in the chiral limit $K\to\pi\pi$ amplitudes can be related to $K\to
\pi$ and $K\to vacuum$ amplitudes, valuable information on the $\Delta I =1/2$
rule can also be obtained 
by evaluating in qcd4 --for several ``small'' external momenta 
and quark mass values-- the (simpler) matrix elements
$\langle\pi^+(p)|\widehat{\cal O}^{\pm}_{VV+AA}
|K^+(q)\rangle $ and $\langle\pi^0(p)|\widehat{\cal
O}^{\pm}_{VV+AA} |K^0(q)\rangle\ $, with the bare four
quark operator given by
\beqn
{\cal O}^{\pm}_{VV+AA} &=& \half
\Big[ (\bar{s}\gamma_\mu u) (\bar{u} \gamma_\mu d) \pm
(\bar{s}\gamma_\mu d) (\bar{u} \gamma_\mu u) \Big]  -  \half
\Big[u
\leftrightarrow c \Big]  +
\nonumber \\[-1pt minus 1pt]&&
+ \half \Big[ (\bar{s}\gamma_\mu\gamma_5 u) (\bar{u}
\gamma_\mu\gamma_5 d) \pm (\bar{s}\gamma_\mu\gamma_5 d) (\bar{u}
\gamma_\mu\gamma_5 u) \Big] - \half \Big[ u \leftrightarrow c \Big] \,
.\qquad
\label{O_pm_VVAA}
\eeqn
In analogy with the case of sect.~(\ref{3B}), one can
extract the bare matrix elements
$\< \pi^+\pi^- | {\cal Q}^{\; \pm}_{VA+AV} | K^0 \>$ and
$\< \pi^0\pi^0 | {\cal Q}^{\; \pm}_{VA+AV} | K^0 \>$
in a 4s10v model
from (connected) correlators of the form~(\ref{GENCORR}), namely
\beq
C_{\pm , K\pi}^{(4s10v)}(x,y) = \langle
\Phi_{\pi}(x) {\cal Q}^{\pm}_{VV+AA}(0) \Phi^\dagger_{K}(y)
\rangle_{\rm conn}^{(4s10v)} \, , \label{KPI_4s10vmod}
\eeq
where $K$ and $\pi$ is either $K^0$ and $\pi^0$ or $K^+$ and $\pi^+$,
the interpolating $K$ and $\pi$ fields involve only the $u$, $d$ and
$s$ valence quarks and, in analogy to eq.~(\ref{Q_01_23_COMB}),
$~{\cal Q}^{\pm}_{VV+AA} = {\cal O}^{\pm \, [0]}_{VV+AA}
 + {\cal O}^{\pm \, [1]}_{VV+AA}
 - {1 \over 2} {\cal O}^{\pm \, [2]}_{VV+AA}
 - {1 \over 2} {\cal O}^{\pm \, [3]}_{VV+AA}~$.
To have no unwanted mixings the lattice regularization
can now be taken such that
\vspace{-0.2cm}
\beq
-r_s = r_d =
r_u = -r_{u'} = r_{u''} =  -r_{u'''} = r_c = -r_{c'} = r_{c''} =
-r_{c'''} \, .
\eeq

\section{Contact with QCD correlators and matrix elements} \label{CONTACT}

Taking the example of the $K^0$--$\bar{K}^0$ mixing amplitude,
we briefly explain the relation between the correlators --and 
derived matrix elements-- in the chosen 4s6v model and those of 
qcd4~\footnote{For more details and the analogous arguments
relevant for the amplitudes discussed in sect.~\ref{3B}--\ref{3C}
see Ref.~\cite{FR2}.}.

%The crucial point is that the 4s6v correlator~$C_{\Phi_\beta {\cal Q} 
%\Phi_\alpha }^{(4s6v)}(x,y)$ (see eq.~(\ref{GENCORR})), with
The crucial point is that the 4s6v correlator~(\ref{GENCORR}), with
$\Phi_\beta = \bar{d}' \gamma_5 s'$, $\Phi_\alpha = \bar{d} \gamma_5 s$
and the operator ${\cal Q}_{VV+AA}^{\Delta S=2}$ specified in
eq.~(\ref{OP_BK}), coincides with the qcd4 correlator~(\ref{CORRBK}),
if (all) the quarks of the same physical flavour are regularized
in the same way in the two theories. The equality immediately follows
from the identity of Wick contractions. In particular it implies that
``corresponding'' local operators, such as ${\cal Q}_{VV+AA}^{\Delta S=2}$
in 4s6v and ${\cal O}_{VV+AA}^{\Delta S=2}$ in qcd4, admit the same
renormalization constant (and thus equal non-perturbative anomalous dimension).
Moreover renormalized (current) quark masses of the same flavour
are taken equal in 4s6v and qcd4. Under these conditions,
the equality of $C_{\Phi_\beta {\cal Q} \Phi_\alpha }^{(4s6v)}(x,y)$
and $C_{\Phi_\beta {\cal O} \Phi_\alpha }^{(qcd4)}(x,y)$ carries over
to the continuum limit, implying equality (as $a \to 0$) of the renormalized 
matrix elements that can be extracted from them, including $\hat{B}_K$.
 
The advantage of working with the 4s6v model rather than with qcd4 is
the greater flexibility in the choice of its lattice regularization,
which results from the possibility of giving different values to the Wilson 
parameters of different replica of the same physical flavour. This flexibility
%which results from the possibility of giving different Wilson parameter
%values to different replica of the same physical flavour. This flexibility
has been exploited in sect.~\ref{3A} by taking 
$~r_d = r_{d'} = r_s = -r_{s'}~$ and is the key to obtain
the discrete symmetries that ensure~\cite{FR2} multiplicative
renormalizability ({\it i.e.} no mixing) of the operator
${\cal Q}_{VV+AA}^{\Delta S=2}$.

%% PIPPO
%identity of qcd4 and 4s6v bare correlators and consequences on Z's \\
%matched quark mass and operator renormalization conditions \\
%identity of qcd4 and 4s6v renormalized correlator ($a \to 0$) \\
%clever choice of the UV regularization of the 4s6v correlator (see above) \\
%O(a) improvement follows from general arguments above, size of O($a^2$)? \\

%%%\section{Conclusions}

\section*{Acknowledgements}
We thank the LOC of Lattice 2005 for the very stimulating atmosphere
of the conference.
% and ... for valuable discussions. 
%Partial support by MIUR-Italy is also acknowledged.                                                                                               

\end{document}